\documentclass[
 preprint,%
 amssymb, amsmath,%
 aip,cha,%
]{revtex4-1}

\usepackage{graphicx}          
\usepackage{graphics} 
\usepackage{epsfig} 
\usepackage{mathptmx} 
\usepackage{times} 
\usepackage{amsmath} 
\usepackage{amssymb}  
\usepackage{floatrow}
\usepackage{enumerate}
\usepackage{xcolor}
\usepackage{mathrsfs}
\usepackage{amsfonts}

\DeclareMathOperator{\Tr}{Tr}

\usepackage{bm}%
\usepackage[colorlinks=true,linkcolor=blue]{hyperref}%
\usepackage[titletoc]{appendix}

\expandafter\ifx\csname package@font\endcsname\relax\else
 \expandafter\expandafter
 \expandafter\usepackage
 \expandafter\expandafter
 \expandafter{\csname package@font\endcsname}%
\fi
\begin{document}

\title{Fault Tolerant Filtering and Fault Detection for Quantum Systems Driven By Fields in Single Photon States}%

\author{Qing Gao}
\email{qing.gao.chance@gmail.com}
\author{Daoyi Dong}
\email{daoyidong@gmail.com}
\author{Ian R. Petersen}
\email{i.r.petersen@gmai.com}
\affiliation{School of Engineering and Information Technology, University of New South Wales, Canberra, ACT, 2600, Australia} 
\author{Herschel Rabitz}%
\email{hrabitz@princeton.edu}
\affiliation{Department of Chemistry, Princeton University, Princeton 08544, New Jersey, USA}%

\date{\today}%

\begin{abstract}
The purpose of this paper is to solve a fault tolerant filtering and fault detection problem for a class of open quantum systems driven by a continuous-mode bosonic input field in single photon states when the systems are subject to stochastic faults. Optimal estimates of both the system observables and the fault process are simultaneously calculated and characterized by a set of coupled recursive quantum stochastic differential equations.
\end{abstract}
\maketitle

\section{Introduction}
Due to the presence of inherent uncertainties in quantum measurements, the theory of quantum filtering plays a fundamental role in quantum measurement based feedback control, which is similar to what an optimal filter does in classical stochastic control systems with partial or noisy observations. Since the publishing of Belavkin's early work \cite{Belavkin1980, Belavkin1988, Belavkin1992} and independent work in the physics community \cite{Carmichael1993}, quantum filtering theory has become routine and is applied in many research areas. For the modern form of quantum filtering, we refer to the work by Bouten \emph{et al.} \cite{Bouten2007}.

In practice, classical randomness may be introduced into the system dynamics of quantum systems, which requires that both classical and quantum randomness should be dealt with simultaneously. For example, the existence of stochastic fluctuations in magnetic flux or gate voltages may cause random changes in the Hamiltonian of a superconducting quantum system \cite{Dong2015}. A spin system may be subject to stochastically fluctuating fields that will introduce classical randomness into the system dynamics \cite{Dong2012}. For an atom interacting with a laser beam, classical randomness arises in the atomic dynamics due to the occurrence of stochastic faults in the laser device \cite{Viola2003, Khodjasteh2005}. For an open quantum system, the introduction of random system Hamiltonians in the system dynamics results in a unitary system evolution that depends on some classical random variables. Consequently, the quantum filter has to be redesigned so that the optimal estimates of system observables can be calculated. In addition, estimation of the fault process is of fundamental significance in applications like guiding control law design.

Single photons are are often used in the realization of all-optical quantum networks and quantum communication protocols. A single photon is a non-classical state of light in the sense that it cannot be described in terms of a classical electric field, and is fundamentally different from a Gaussian state. This situation makes single photons very useful in quantum information processing. Recently some results have been reported on the analysis of quantum systems driven by a single photon input. For instance, the interaction between single photon packets and excited atoms is analyzed in \cite{Elyutin2012}; in \cite{Zhang2013}, detailed results of the response of a linear quantum system to single photon input fields was given; quantum filtering of open quantum systems driven by fields in single photon states can be found in \cite{Koshino, Gough2012a, Gough2012, Gough2013}. In this paper, we concentrate on a class of open quantum systems probed by a continuous-mode bosonic input field in single photon states and subject to stochastic faults.  By applying a quantum-classical conditional expectation method in our recent work \cite{Gao2015}, a fault tolerant design of the quantum filter for this class of open quantum systems is given. The equations of the conditional density distribution of the fault process are also obtained. Using this result, a possible criteria for fault detection is provided.

\section{Heisenberg-Picture Dynamical Models}
The following typical experimental setup in quantum optics is considered in this paper: a laser probe field interacts with a cloud of atoms trapped in a cavity and is subsequently continuously detected by a homodyne detector which gives rise to a classical measurement signal. In particular, we consider the case where the input field is placed in a continuous-mode single photon state $\left| 1_{\xi}\right>$ \cite{Gough2012}. Here $\xi(t)$ is a normalized complex-valued function representing the single photon wave packet shape and satisfying $\int_0^{\infty} |\xi(\tau)|^2d\tau=1$. To specify a light field with non-Gaussian statistics, e.g., a field in a single photon state, involves in principle the specification of all possible correlation functions and is far from being practical. One possible way of specifying such non-Gaussian statistics can be achieved by modelling the apparatus that produces the light, and coupling the generated output light from the obtained model into the quantum system under study. Following a similar idea in \cite{Gough2012}, where an ancilla two-level quantum system $A_s$ driven by a vacuum field was used to model the effect of the single photon state for $B(t)$ on the atom system $G_s=(I, L_{G_s}, H_{G_s})$, we start from the cascaded system as in Fig. 1. The ancilla system $A_s$ is initially prepared in its excited state $\left|\uparrow\right>$ and its interaction with the vacuum input field is described by 
\begin{equation}
(S_{A_s}, L_{A_s}, H_{A_s})=(I, \lambda(t) \sigma_-,0),\label{gquantum1}
\end{equation} 
where $\lambda(t)=\frac{\xi(t)}{\sqrt{\omega(t)}}$ and $\omega(t)=\int_t^{\infty} |\xi(\tau)|^2d\tau$. The output field of the ancilla system $A_s$ is then fed into the atom system $G_s$, i.e., $A_s$ and $G_s$ form a cascaded quantum network driven by a vacuum field. The unitary $U(t)$ of the cascaded system satisfies the following quantum stochastic differential equation:
 \begin{eqnarray}
dU(t)&=&\left\{\left(-iH-\frac{1}{2}L^{\dagger}L\right)dt+LdB^{\dagger}(t)-L^{\dagger}dB(t)\right\}U(t), \label{gquantum2}
\end{eqnarray}
with initial condition $U(0)=I$. We have assumed $\hbar=1$ by using atomic units in this paper. According to quantum network theory \cite{Gough2009, Gardiner2000} the bounded Hermitian operator $H=H_{G_s}+H_{A_s}+\Im\{L_{G_s}^{\dagger}L_{A_s}\}$, where $\Im\{X\}$ is the imaginary part of $X$, is the Hamiltonian of the cascaded system. $B(t)$ and $B^{\dagger}(t)$ are the field operators representing quantum noises, and together with the cascaded system operator $L=L_{A_s}+L_{G_s}$ they model the interaction between the cascaded system and the laser probe field. From quantum stochastic calculus \cite{Hudson1984, Holevo1991,Parthasarathy1992}, the forward difference of quantum noise $dB(t)=B(t+dt)-B(t)$ satisfies
\begin{eqnarray*}
&&dB(t)dB^{\dagger}(t)=dt,\\
&&dB^{\dagger}(t)dB(t)=dB(t)dB(t)=dB^{\dagger}(t)dB^{\dagger}(t)=0.
\end{eqnarray*}
In terms of system state, letting system $G_s$ be initialized in $\pi_0$, we write $\rho_0=\left|\uparrow\right>\left<\uparrow\right|\otimes\pi_0\otimes \left|\upsilon \right>\left<\upsilon \right|$, where $\left|\upsilon \right>$ represents the vacuum state. It is noted that (\ref{gquantum2}) is given in $It\hat o$ form, as will all stochastic differential equations in this paper.

\begin{figure}\label{Fig. 1}
\centering
\includegraphics[width=5in]{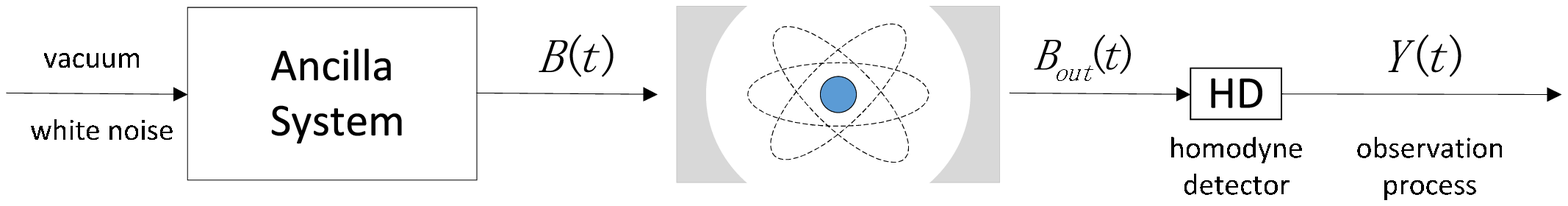}
\caption{A schematic representation of the physical scenario, where an ancilla system is used to model the effect of the single photon state for $B(t)$ on the atom system.}
\end{figure}

The cascaded system and the laser field form a composite system on the Hilbert space $\mathcal{C}=\mathbb{C}^2\otimes \mathcal{H}_{\mathcal{S}}\otimes \mathcal{E}=\mathbb{C}^2\otimes \mathcal{H}_{\mathcal{S}}\otimes \mathcal{E}_{t]}\otimes \mathcal{E}_{(t}$, where we have used the continuous temporal tensor product decomposition of the Fock space $\mathcal{E}=\mathcal{E}_{t]}\otimes \mathcal{E}_{(t}$ into the past and future components. In what follows, we assume that $\dim(\mathcal{H}_{\mathcal{S}})=n<\infty$. The observables of the ancilla system and the atom system are described by self-adjoint operators on $\mathbb{C}^2$ and $\mathcal{H}_{\mathcal{S}}$, respectively. Any cascaded system operator $A\otimes X$ at time $t$ is given by $j_t(A\otimes X)=U^{\dagger}(t)(A\otimes X\otimes I)U(t)$ and satisfies \cite{Gardiner2000, Breuer2002}
 \begin{eqnarray}
dj_t(A\otimes X)=j_t(\mathscr{L}_{L,H}(A\otimes X))dt+j_t([A\otimes X,L])dB^{\dagger}(t)+j_t([L^{\dagger}, A\otimes X])dB(t), \label{gquantum3}
\end{eqnarray}
where the so-called Lindblad generator is defined by 
\begin{eqnarray*}
\mathscr{L}_{L,H}(X)=-i[X,H]+L^{\dagger}XL-\frac{1}{2}(L^{\dagger}LX+XL^{\dagger}L).
\end{eqnarray*}

The above description is only an idealization of the real physical interactions. In many practical implementations, the system Hamiltonian may be randomly changing because of, e.g., the introduction of faulty control Hamiltonians when implementing an imperfect experimental setup \cite{Viola2003, Khodjasteh2005}, or random fluctuations of the external electromagnetic field (laser intensity) \cite{Ruschhaupt2012, Dong2015}. In this case, the system Hamiltonian can be appropriately modelled by a time-varying random Hermitian operator functional $H(F(t))$ that depends on some classical stochastic process $F(t)$. Using the quantum $It \hat o$ rule \cite{Hudson1984}, one still has $d(U^{\dagger}(t)U(t))=d(U(t)U^{\dagger}(t))=0$ in (\ref{gquantum2}), which implies that $U(t)$ is a \emph{unitary operator} depending on the stochastic process $F(t)$. For the sake of simplicity, we still write $U(t)$ instead of the functional form $U(F,t)$. From the unitarity of $U(t)$, one can conclude that the commutativity of operators is preserved, that is, $[j_t(A), j_t(B)]=0$ if $[A, B]=0$ where $A,B$ are two cascaded system operators on $\mathcal{C}$. Here the commutator is defined by $[A, B]=AB-BA$.

It is observed from the stochastic model  (\ref{gquantum2}) that $U(t)$ depends on $B(t')$ and $B^{\dagger}(t')$, $0\leq t'< t$, since the increment operators $dB(t)$ and $dB^{\dagger}(t)$ are future pointing \cite{Gardiner2000}. Consequently,
\begin{equation}
[U(t), dB(t)]=[U(t), dB^{\dagger}(t)]=0. \label{gquantum4}
\end{equation}
Similarly, the time evolution operator $U(t,s)=U(t)U^{\dagger}(s)$ from time $s$ to time $t$ depends only on the field operators $dB(s')$ and $dB^{\dagger}(s')$ with $s'$ between $s$ and $t$. The commutation relations then show that
\begin{equation}
[U(t,s), B(\tau)]=[U(t,s), B^{\dagger}(\tau)]=0, \tau \leq s. \label{gquantum5}
\end{equation}

In a quantum optical system, the measurement of a system observable is usually performed by detecting the probe field observables, aiming to not perturb the subsequent evolution of the system observable. This is the basic concept behind so-called quantum non-demolition (QND) measurements, which is adopted in this work. For the physical scenario under consideration in this paper, the observation process is given by $Y(t)=j_t(Q(t))=U^{\dagger}(t)(I\otimes I\otimes Q(t))U(t)$ where $Q(t)=B(t)+B^{\dagger}(t)$ is the real quadrature of the input field satisfying $[Q(t), Q(s)]=0$. Physically, $Y(t)$ may represent the integrated photocurrent arising in a perfect homodyne photon detection setup. Combing (\ref{gquantum4}) and (\ref{gquantum5}) with the fact that $[I\otimes I\otimes Q(t), A\otimes X\otimes I]=0$, it is easy to show that: (i) $[Y(t), Y(s)]=0$ at all times $s, t$ and (ii) $[Y(s), j_t(A\otimes X)]=0, \forall s\leq t$. These two properties guarantee that (i) $Y(t)$ can be continuously monitored without perturbing the subsequent system evolution, and (ii) it is possible to make a conditional statistical inference of any observable $j_t(A\otimes X)$ from the history of $Y(t)$. In addition, by using the quantum $It\hat o$ rule, one has 
\begin{equation}
dY(t)=U^{\dagger}(t)(L+L^{\dagger})U(t)dt+dQ(t), \label{gquantum6}
\end{equation}
from which $Y(t)$ has the form of $j_t(L+L^{\dagger})=U^{\dagger}(t)(L+L^{\dagger})U(t)$ with a noise term $Q(t)$.

\section{Quantum Filtering and Fault Detection of Quantum Systems Driven by Single Photon Fields.}
In classical (non-quantum) engineering, a fault (abrupt or incipient) refers to any kind of undesired deviation of the characteristic properties or parameters of the system from normal conditions, which can often lead to a reduction in performance or even loss of key functions in the physical plant. Thus a fault tolerant design possesses practical significance in engineering. Recall the class of quantum systems described in Section II. In the laser-atom interaction picture, the spectrum of the classical electromagnetic field enclosed in a cavity depends on the geometric construction of the cavity, while the laser-atom interaction is described by a dipole interaction Hamiltonian that depends on the intensity of the electromagnetic field \cite{Walther2006}. Therefore, if the setup of the cavity suffers from a fault, e.g., an abrupt variation in its geometry, the intensity of the electromagnetic field inside will be unavoidably changed and an unexpected additional Hamiltonian term will be introduced into the quantum system. In this case, the atom system Hamiltonian will be given by a random Hermitian operator $H_{G_s}(F(t))$ that depends on the fault process $F(t)$, and the cascaded system Hamiltonian is then given by $H(F(t))=H_{G_s}(F(t))+\Im\{L_{G_s}^{\dagger}\lambda(t) \sigma_-\}$.

In practice, the fault process is normally modelled on a classical probability space $(\Omega,\mathcal{F},\mathcal{P})$ by a continuous-time Markov chain $\{F(t)\}_{t\geq 0}$ adapted to $\{\mathcal{F}_t\}_{t\geq 0}$ \cite{Davis1975, Hibey1999, Elliott1995}, which coincides with the phenomenon that physical systems may transit among a set of different faulty modes at random time points. The state space of $F(t)$ is often chosen to be the finite set $\mathbb{S}=\{e_1, e_2,...,e_N\}$ (for some positive integer $N$) of canonical unit vectors in $\mathbb{R}^N$. Let $p_t=(p_t^1, p_t^2,...,p_t^N)^T$ be the probability distribution of $F(t)$, i.e., $p_t^k=\mathcal{P}(F(t)=e_k), k=1,2,...,N$ and suppose that the Markov process $F(t)$ has a so-called Q matrix or transition rate matrix $\Pi=(a_{jk})\in \mathbb{R}^{N\times N}$. Then $p_t$ satisfies the forward Kolmogorov equation $ \frac{dp_t}{dt}=\Pi p_t$. Because $\Pi$ is a Q matrix, we have $a_{jj}=-\sum_{j\neq k}a_{jk}$, and $a_{jk}\geq 0, j\neq k$. Then $F(t)$ is a corlol process \cite{Elliott1995} satisfying the following stochastic differential equation:
\begin{equation}
dF(t)=\Pi F(t)dt+dM(t), \label{gquantum13}
\end{equation}
where $M(t)=F(t)-F(0)-\int_0^t\Pi F(\tau^-)d\tau$ is an $\{\mathcal{F}_t\}$ martingale \cite{Elliott1995} that satisfies
$\sup \limits_{0\leq t \leq T} \mathbb{E}(|M(t)|^2)< \infty.$ In addition, we assume that the statistics of $F_l$ are unperturbed by quantum measurements due to the lack of significant quantum backaction on classical systems.

\textcolor{blue}{
\textbf{Example 3.1.}  Examples of the above description of a faulty mode in open quantum systems can be found in literature. For example, in \cite{Khodjasteh2005}, the quantum system may have a ``rectangular'' pulsive (piecewise constant) external Hamiltonian which was supposed to be bounded and applied to the quantum system at time $T$. In practice, it is reasonable to assume that $T$ is an exponential random variable with a constant parameter $\lambda>0$. Let
\begin{equation}
z_t=\left\{
\begin{array}{cc}
0, \mbox{ if } t<T;\\
1, \mbox{ if } t\geq T.
\end{array}\right. \label{add1}
\end{equation}
Then the system Hamiltonian has the form of $H(t)=H_0+z_tH_1$ where $H_0$ represents the free Hamiltonian and $H_1$ is the external Hamiltonian. From (\ref{add1}) one knows $z_t$ is a Poisson type stochastic process with rate $\lambda$ stopped at its first jump time $T$ and has an associated martingale $M_t=z_t-\lambda \min(t,T)$. Thus we have
\begin{equation}
dz_t=\lambda(1-z_t)dt+dM_t. \label{add2}
\end{equation}}
\vspace{-0.1mm}
\textcolor{blue}{
Then the fault process $F(t)=[1-z_t,z_t]'$ takes values in $\{e_1,e_2\}$ and satisfies
\begin{equation}
dF(t)=\left[
\begin{array}{cc}
-\lambda & 0\\
\lambda  & 0
\end{array}
\right]F(t)+\left[
\begin{array}{c}
-1\\
1
\end{array}
\right]dM_t. \label{add3}
\end{equation}}

The aim of this work is to derive the equations of the fault tolerant quantum filter and fault detection for this class of open quantum systems driven by fields in single photon states. To be specific, we use a reference probability approach to simultaneously find the least-mean-square estimates of a system observable $X\in \mathscr{B}(\mathcal{H})$ at time $t$ and the fault process $F(t)$ for the quantum system under consideration, given the observation process $Y(t)$. \textcolor{blue}{Because now we have both quantum and classical randomnesses to be dealt with, we introduce a combined quantum-classical expectation operator $\tilde{\mathbb{P}}(\cdot)=\mathbb{E}_{\mathcal{P}}\{\Tr\{\rho_0 (\cdot)\}\}: \mathscr{B}(\mathcal{H})\to \mathbb{R}$ to allow for convenient calculations. Then the goal of this work can be accomplished if we can compute the following estimates
\begin{equation}
\sigma_t^j(A\otimes X)=\tilde{\mathbb{P}}(\left<F(t),e_j\right>U^{\dagger}(t)(A\otimes X)U(t)|\mathscr{Y}_t),\label{gquantum14}
\end{equation}
where $A$ is any observable of the ancilla system, $\mathscr{Y}_t$ is the commutative von Neumann algebra generated by $Y(s)$ up to time $t$, and $\left<\cdot, \cdot\right>$ is the inner product in $\mathbb{R}^N$. The definition of the quantum-classical conditional expectation can be found in Appendix and a complete treatment can be found in \cite{Gao2015}. This conditional expectation can be formulated using a framework describing random observables in \cite{Bouten2009} and in \cite{Gao2015} it has been explained under this framework. In fact, a set of commutative random observables is isomorphic to a set of classical random variables on a unique classical probability space model, which implies that the joint statistics between a set of commutative random observables can be well defined using associated classical concepts. Generally, the quantum-classical conditional expectation is equivalent to a particular quantum conditional expectation \cite{Bouten2007, Bouten2009} and contains the classical conditional expectation \cite{Bertsekas2002} as a special case. The elementary properties of classical conditional expectation, for example, linearity, positivity, the tower property and ``taking out what is known'' \cite{Bertsekas2002}, still hold for the above defined conditional expectation. In addition, we have
\begin{equation}
\|\left<F(t),e_j\right>U^{\dagger}(t)(A\otimes X)U(t)-\sigma_t^j(A\otimes X)\|_{\tilde{\mathbb{P}}}\leq \|\left<F(t),e_j\right>U^{\dagger}(t)(A\otimes X)U(t)-Y\|_{\tilde{\mathbb{P}}},
\end{equation}
for all $Y\in \mathscr{Y}_t$, where $\|X\|_{\tilde{\mathbb{P}}}=\tilde{\mathbb{P}}(X^{\dagger}X)$. This guarantees the optimality of $\sigma_t^j(A\otimes X)$ in the mean square sense.}

The following lemma plays a fundamental role in deriving the quantum filtering equation and the fault detection equation.

\textbf{Lemma 3.1.} Let $V(t)$ be a random operator satisfying the quantum stochastic differential equation
\begin{equation}
dV(t)=\left\{\left(-iH(F(t))-\frac{1}{2}L^{\dagger}L\right)dt+LdQ(t)\right\}V(t),\label{gquantum15}
\end{equation}
with $V(0)=I$. Then
\begin{equation}
\sigma_t^j(A\otimes X)=U^{\dagger}(t)\frac{\tilde{\mathbb{P}}\left(\left<F(t),e_j\right>V^{\dagger}(t)(A\otimes X)V(t)|\mathscr{Q}_t\right)}{\tilde{\mathbb{P}}(V^{\dagger}(t)V(t))|\mathscr{Q}_t)}U(t), \label{gquantum16}
\end{equation}
where $\mathscr{Q}_t$ is the commutative von Neumann algebra generated by $Q(s)$ up to time $t$.

\emph{Proof.} See the Appendix.

Based on Lemma 3.1, the following theorem can be obtained.

\textbf{Theorem 3.1.} The conditional expectation $\sigma_t^j(A\otimes X)$ satisfies the following quantum stochastic differential equation:
\begin{eqnarray}
d\sigma_t^j(A\otimes X)&=&\left(\sum_{k=1}^Na_{jk}\sigma_t^k(A\otimes X)+\sigma_t^j(\mathscr{L}_{L, H(e_j)}(A\otimes X))\right)dt\nonumber\\
&&+\left(\sigma_t^j((A\otimes X)L+L^{\dagger}(A\otimes X))-\sigma_t^j(A\otimes X)\sum_{k=1}^N\sigma_t^k(L+L^{\dagger})\right)dW(t), \label{gquantum17}
\end{eqnarray}
where the so-called Lindblad generator is given by
\begin{equation*}
\mathscr{L}_{L, H}(X)=i[H,X]+L^{\dagger}XL-\frac{1}{2}(L^{\dagger}LX+XL^{\dagger}L),
\end{equation*}
and the innovation process $W(t)=Y(t)-\int_0^t \sum_{k=1}^N\sigma_s^k(L+L^{\dagger})ds$ is a Wiener process under $\tilde{\mathbb{P}}$.

\emph{Proof.} Using the $It\hat o$ product rule, and from the mutual independence of $\{Q(t), M(t), F(0)\}$, the following result can be obtained:
\begin{eqnarray}
&&\tilde{\mathbb{P}}(\left<F(t),e_j\right>V^{\dagger}(t)(A\otimes X)V(t)|\mathscr{Q}_t)\nonumber\\
&=&\tilde{\mathbb{P}}(\left<F(0),e_j\right>X)+\int_0^t \tilde{\mathbb{P}}(\left<\Pi F(s),e_j\right>V^{\dagger}(s)XV(s)|\mathscr{Q}_s)ds\nonumber\\
&&+\int_0^t \tilde{\mathbb{P}}\left(\left<F(s),e_j\right>V^{\dagger}(s)\mathscr{L}_{L, H(e_j)}(X)V(s)|\mathscr{Q}_s\right)ds\nonumber
\end{eqnarray}
\begin{eqnarray}
&&+\int_0^t \tilde{\mathbb{P}}\left( \left<F(s),e_j\right>V^{\dagger}(s)(XL+L^{\dagger}X)V(s)|\mathscr{Q}_s\right)dQ(s).  \label{gquantum18}
\end{eqnarray}

In addition, from the property of the Q matrix $\Pi$, we have $\left<\Pi F(s),e_j\right>=\left<F(s),\Pi^{T}e_j\right>=\left<F(s),\sum_{k=1}^N a_{jk}e_k\right>=\sum_{k=1}^N a_{jk}\left<F(s),e_k\right>.$ Define $h_t^j(A\otimes X)=\tilde{\mathbb{P}}(\left<F(t),e_j\right>V^{\dagger}(t)(A\otimes X)V(t)|\mathscr{Q}_t)$. Then from (\ref{gquantum18}) we have
\begin{eqnarray}
dh_t^j(X)=\left(\sum_{k=1}^N a_{jk} h_t^k(X)+h_t^j(\mathscr{L}_{L, H(e_j)}(X))\right)dt+h_t^j(XL+L^{\dagger}X)dQ(t).\label{gquantum19}
\end{eqnarray}
(\ref{gquantum17}) can be obtained from (\ref{gquantum19}) using some manipulations in quantum stochastic calculus. The proof is thus completed.

We are interested in the conditional estimation of the system operator only. Using quantum $It\hat o$ rule, one has
\begin{eqnarray}
\left(
\begin{array}{cc}
 \mathscr{L}_{L_{A_s}, H_{A_s}}(I)  & \mathscr{L}_{L_{A_s}, H_{A_s}}(\sigma_{-})\\
  \mathscr{L}_{L_{A_s}, H_{A_s}}(\sigma_{+}) &  \mathscr{L}_{L_{A_s}, H_{A_s}}(\sigma_{+}\sigma_{-}) 
\end{array}
\right)=\left(
\begin{array}{cc}
 0  & -\frac{|\xi(t)|^2}{2\omega(t)}\sigma_{-}\\
  -\frac{|\xi(t)|^2}{2\omega(t)}\sigma_{+} &  -\frac{|\xi(t)|^2}{\omega(t)}\sigma_{+}\sigma_{-}
\end{array}
\right).
\end{eqnarray}
Then by defining the following conditional expectations:
\begin{eqnarray}
\left(
\begin{array}{cc}
 \sigma_{t,00}^j(X)  & \sigma_{t,01}^j(X) \\
  \sigma_{t,10}^j(X) &   \sigma_{t,11}^j(X)  
\end{array}
\right)=
\left(
\begin{array}{cc}
  \frac{\sigma_t^j((\sigma_{+}\sigma_{-})\otimes X)}{\omega(t)}   &  \frac{\sigma_t^j(\sigma_{+}\otimes X)}{\sqrt{\omega(t)}} \\
   \frac{\sigma_t^j(\sigma_{-}\otimes X)}{\sqrt{\omega(t)}} &   \sigma_t^j(I\otimes X)   
\end{array}
\right), \label{gquantum20}
\end{eqnarray}
the following coupled nonlinear stochastic differential equation can be obtained from (\ref{gquantum17}).
\begin{eqnarray}
d\sigma_{t,11}^j(X)&=&\left\{\sum_{k=1}^Na_{jk}\sigma_{t,11}^k(X)+\sigma_{t,11}^j(\mathscr{L}_{L_{G_s}, H_{G_s}(e_j)}(X))+\sigma_{t,01}^j([X,L_{G_s}])\xi^{\dagger}(t)+\sigma_{t,10}^j([L_{G_s}^{\dagger},X])\xi(t)\right\}dt\nonumber\\
&&+\left\{\sigma_{t,11}^j(XL_{G_s}+L_{G_s}^{\dagger}X)+\sigma_{t,01}^j(X)\xi^{\dagger}(t)+\sigma_{t,10}^j(X)\xi(t)-\sigma_{t,11}^j(X)K_t\right\}dW(t)\nonumber\\
d\sigma_{t,10}^j(X)&=&\left\{\sum_{k=1}^Na_{jk}\sigma_{t,10}^k(X)+\sigma_{t,10}^j(\mathscr{L}_{L_{G_s}, H_{G_s}(e_j)}(X))+\sigma_{t,00}^j([X,L_{G_s}])\xi^{\dagger}(t)\right\}dt\nonumber\\
&&+\left\{\sigma_{t,10}^j(XL_{G_s}+L_{G_s}^{\dagger}X)+\sigma_{t,00}^j(X)\xi^{\dagger}(t)-\sigma_{t,10}^j(X)K_t\right\}dW(t)\nonumber\\
d\sigma_{t,01}^j(X)&=&\left\{\sum_{k=1}^Na_{jk}\sigma_{t,01}^k(X)+\sigma_{t,01}^j(\mathscr{L}_{L_{G_s}, H_{G_s}(e_j)}(X))+\sigma_{t,00}^j([X,L_{G_s}]\xi^{\dagger}(t))\right\}dt\nonumber\\
&&+\left\{\sigma_{t,01}^j(XL_{G_s}+L_{G_s}^{\dagger}X)+\sigma_{t,00}^j(X)\xi^{\dagger}(t)-\sigma_{t,01}^j(X)K_t\right\}dW(t)\nonumber
\end{eqnarray}
\begin{eqnarray}
d\sigma_{t,00}^j(X)&=&\left\{\sum_{k=1}^Na_{jk}\sigma_{t,00}^k(X)+\sigma_{t,00}^j(\mathscr{L}_{L_{G_s}, H_{G_s}(e_j)}(X))\right\}dt\nonumber\\
&&+\left\{\sigma_{t,00}^j(XL_{G_s}+L_{G_s}^{\dagger}X)-\sigma_{t,00}^j(X)K_t\right\}dW(t) \label{gquantum21}
\end{eqnarray}
where $K_t=\sigma_{t,11}^j(L_{G_s}+L_{G_s}^{\dagger})+\sigma_{t,01}^j(I)\xi(t)+\sigma_{t,01}^j(I)\xi^{\dagger}(t)$ and the innovation process $W(t)$ is given by $W(t)=dY(t)-K_tdt$.

With the following relations:
\begin{eqnarray}
\left(
\begin{array}{cc}
 \sigma_{t,00}(X)  & \sigma_{t,01}(X) \\
  \sigma_{t,10}(X) &   \sigma_{t,11}(X)  
\end{array}
\right)&=&
\left(
\begin{array}{cc}
  \frac{\sum_{k=1}^N \sigma_t^k((\sigma_{+}\sigma_{-})\otimes X)}{\omega(t)}   &  \frac{\sum_{k=1}^N \sigma_t^k(\sigma_{+}\otimes X)}{\sqrt{\omega(t)}} \\
   \frac{\sum_{k=1}^N \sigma_t^k(\sigma_{-}\otimes X)}{\sqrt{\omega(t)}} &   \sum_{k=1}^N\sigma_t^k(I\otimes X)   
\end{array}
\right)\nonumber\\
&=&\left(
\begin{array}{cc}
 \tilde{\mathbb{P}}(U^{\dagger}(t)((\sigma_{+}\sigma_{-})\otimes X)U(t)|\mathscr{Y}_t) &  \frac{\tilde{\mathbb{P}}(U^{\dagger}(t)(\sigma_{+}\otimes X)U(t)|\mathscr{Y}_t)}{\sqrt{\omega(t)}}\\
 \frac{\tilde{\mathbb{P}}(U^{\dagger}(t)(\sigma_{-}\otimes X)U(t)|\mathscr{Y}_t)}{\sqrt{\omega(t)}} & \tilde{\mathbb{P}}(U^{\dagger}(t)(I\otimes X)U(t)|\mathscr{Y}_t)
 \end{array}
 \right), \hspace{3mm} \label{gquantum22}
\end{eqnarray}
the following coupled nonlinear stochastic differential equations can be obtained
\begin{eqnarray}
d\sigma_{t,11}(X)&=&\left\{\sum_{k=1}^N\sigma_{t,11}^k(\mathscr{L}_{L_{G_s}, H_{G_s}(e_k)}(X))+\sigma_{t,01}([X,L_{G_s}])\xi^{\dagger}(t)+\sigma_{t,10}([L_{G_s}^{\dagger},X])\xi(t)\right\}dt\nonumber\\
&&+\left\{\sigma_{t,11}(XL_{G_s}+L_{G_s}^{\dagger}X)+\sigma_{t,01}(X)\xi^{\dagger}(t)+\sigma_{t,10}(X)\xi(t)-\sigma_{t,11}(X)K_t\right\}dW(t)\nonumber\\
d\sigma_{t,10}(X)&=&\left\{\sum_{k=1}^N\sigma_{t,10}^k(\mathscr{L}_{L_{G_s}, H_{G_s}(e_k)}(X))+\sigma_{t,00}([X,L_{G_s}])\xi^{\dagger}(t)\right\}dt\nonumber\\
&&+\left\{\sigma_{t,10}(XL_{G_s}+L_{G_s}^{\dagger}X)+\sigma_{t,00}(X)\xi^{\dagger}(t)-\sigma_{t,10}(X)K_t\right\}dW(t)\nonumber\\
d\sigma_{t,01}(X)&=&\left\{\sum_{k=1}^N\sigma_{t,01}^k(\mathscr{L}_{L_{G_s}, H_{G_s}(e_k)}(X))+\sigma_{t,00}([X,L_{G_s}]\xi^{\dagger}(t))\right\}dt\nonumber\\
&&+\left\{\sigma_{t,01}(XL_{G_s}+L_{G_s}^{\dagger}X)+\sigma_{t,00}(X)\xi^{\dagger}(t)-\sigma_{t,01}(X)K_t\right\}dW(t)\nonumber\\
d\sigma_{t,00}(X)&=&\sum_{k=1}^N\sigma_{t,00}^k(\mathscr{L}_{L_{G_s}, H_{G_s}(e_k)}(X))dt+\left\{\sigma_{t,00}(XL_{G_s}+L_{G_s}^{\dagger}X)-\sigma_{t,00}(X)K_t\right\}dW(t).\label{gquantum23}
\end{eqnarray}

Note $\sigma_{t,11}(X) =\tilde{\mathbb{P}}(U^{\dagger}(t)(I\otimes X)U(t)|\mathscr{Y}_t)$ is exactly the least-mean-square estimate of the atom observable $X$ at time $t$ and (\ref{gquantum23}) are the \emph{fault tolerant single photon quantum filter equations}. When $\pi_{jk}=0, \forall j\neq k$, $H_{G_s}(F(t))\equiv H_{G_s}$, and this system is partly decoupled and reduces to the single photon quantum filtering equation of $U^{\dagger}(t)XU(t)$ given $\mathscr{Y}_t$ in \cite{Gough2012}:
\textcolor{blue}{
\begin{eqnarray}
d\bar \sigma_{t,11}(X)&=&\left\{\bar \sigma_{t,11}(\mathscr{L}_{L_{G_s}, H_{G_s}}(X))+\bar \sigma_{t,01}([X,L_{G_s}])\xi^{\dagger}(t)+\bar \sigma_{t,10}([L_{G_s}^{\dagger},X])\xi(t)\right\}dt\nonumber\\
&&+\left\{\bar \sigma_{t,11}(XL_{G_s}+L_{G_s}^{\dagger}X)+\bar \sigma_{t,01}(X)\xi^{\dagger}(t)+\bar \sigma_{t,10}(X)\xi(t)-\bar \sigma_{t,11}(X)K_t\right\}dW(t)\nonumber\\
d\bar \sigma_{t,10}(X)&=&\left\{\bar \sigma_{t,10}(\mathscr{L}_{L_{G_s}, H_{G_s}}(X))+\bar \sigma_{t,00}([X,L_{G_s}])\xi^{\dagger}(t)\right\}dt\nonumber\\
&&+\left\{\bar \sigma_{t,10}(XL_{G_s}+L_{G_s}^{\dagger}X)+\bar \sigma_{t,00}(X)\xi^{\dagger}(t)-\bar \sigma_{t,10}(X)K_t\right\}dW(t)\nonumber\\
d\bar \sigma_{t,01}(X)&=&\left\{\bar \sigma_{t,01}(\mathscr{L}_{L_{G_s}, H_{G_s}}(X))+\bar \sigma_{t,00}([X,L_{G_s}]\xi^{\dagger}(t))\right\}dt\nonumber\\
&&+\left\{\bar \sigma_{t,01}(XL_{G_s}+L_{G_s}^{\dagger}X)+\bar \sigma_{t,00}(X)\xi^{\dagger}(t)-\bar \sigma_{t,01}(X)K_t\right\}dW(t)\nonumber\\
d\bar \sigma_{t,00}(X)&=&\bar \sigma_{t,00}(\mathscr{L}_{L_{G_s}, H_{G_s}}(X))dt+\left\{\bar \sigma_{t,00}(XL_{G_s}+L_{G_s}^{\dagger}X)-\bar \sigma_{t,00}(X)K_t\right\}dW(t),\label{revis1}
\end{eqnarray}
where 
\begin{eqnarray*}
\left(
\begin{array}{cc}
 \bar \sigma_{t,00}(X)  & \bar \sigma_{t,01}(X) \\
  \bar \sigma_{t,10}(X) &   \bar \sigma_{t,11}(X)  
\end{array}
\right)=\left(
\begin{array}{cc}
\mathbb{P}(U^{\dagger}(t)((\sigma_{+}\sigma_{-})\otimes X)U(t)|\mathscr{Y}_t) &  \frac{\mathbb{P}(U^{\dagger}(t)(\sigma_{+}\otimes X)U(t)|\mathscr{Y}_t)}{\sqrt{\omega(t)}}\\
 \frac{\mathbb{P}(U^{\dagger}(t)(\sigma_{-}\otimes X)U(t)|\mathscr{Y}_t)}{\sqrt{\omega(t)}} & \mathbb{P}(U^{\dagger}(t)(I\otimes X)U(t)|\mathscr{Y}_t)
 \end{array}
 \right).
\end{eqnarray*}}

In addition, the conditional probability densities of the fault process are given by
\begin{equation}
\hat p_t^j=\mathcal{P}(F(t)=e_j|\mathscr{Y}_t)=\tilde{\mathbb{P}}(\left<F(t),e_j\right>|\mathscr{Y}_t)=\sigma_t^j(I\otimes I), \label{gquantum24}
\end{equation}
which satisfy the following coupled equations using Theorem 3.1:
\begin{eqnarray}
d\hat p_t^j&=&\sum_{k=1}^Na_{jk}\hat p_t^kdt+\left(\sigma_t^j(L+L^{\dagger})-\hat p_t^j\sum_{k=1}^N\sigma_t^k(L+L^{\dagger})\right)dW(t).  \label{gquantum25}
\end{eqnarray}
Let $\hat p_t=[\hat p_t^1,...,\hat p_t^N]'$. Then (\ref{gquantum25}) can be rewritten in a vector form as
\begin{equation}
d\hat p_t=\Pi\hat p_tdt+G(t)dW(t),\label{gquantum26}
\end{equation}
where $G(t)=\sum_{k=1}^Ne_k\sigma_t^k(L+L^{\dagger})-\hat p_t\sum_{k=1}^N\sigma_t^k(L+L^{\dagger})$.
Equation (\ref{gquantum26}) is the corresponding \emph{fault detection equation}. \textcolor{blue}{Here $\sigma_t^j(L+L^{\dagger})=\tilde{\mathbb{P}}(\left<F(t),e_j\right>U^{\dagger}(t)(I\otimes (L+L^{\dagger}))U(t)|\mathscr{Y}_t)$ can be calculated using the recursive stochastic differential equation (\ref{gquantum17}). }

\textcolor{blue}{One can observe that after the classical measurement results are obtained, our knowledge about the probability distribution of the stochastic process $F(t)$ has been refined from the forward Kolmogorov equation $\frac{dp_t}{dt}=\Pi p_t$ to (\ref{gquantum26}). The system of coupled equations (\ref{gquantum25}) or the vector form (\ref{gquantum26}) represents the conditional probability distribution of the system under any faulty mode. It can be used to determine whether a particular type of fault has happened within the system at time $t$. A possible criteria for fault detection is given by
\begin{equation}
\mbox{The $j$th fault happens, if }\hat p_t^j\geq p_0, \label{gquantum27}
\end{equation}
where $1\geq p_0>0$ is a threshold value chosen by the users. Here ``the $j$th fault happens'' means that one could determine at time $t$ that the system Hamiltonian has jumped to $H_{G_s}(e_j)$, which has practical significance in fault repair and fault tolerant control law design. Note that sometimes multiple faulty modes might be determined from the fault detection strategy in (\ref{gquantum27}). To solve this problem one could carefully choose the threshold probability value $p_0$ or using different fault detection criteria, see e.g., \cite{Isermann2006}.}


\section*{Acknowledgement}

This work has been supported by the Australian Research Council (DP130101658, FL110100020).  Herschel Rabitz acknowledges support from the United States NSF (CHE-1058644). The authors acknowledge helpful discussions with Dr. Hendra Nurdin and Prof. Matthew James.

\section*{Appendix}

\textcolor{blue}{\textbf{Definition A1.} (Quantum-classical conditional expectation) \cite{Gao2015} Let $\mathscr{C}$ be a commutative von Neumann algebra on $\mathscr{H}$. Given a $\mathbb{R}^{n_r}$ valued classical random variable $R$ on $(\Omega,\mathcal{F},\mathcal{P})$ and a corresponding unitary operator $U_R$, define $\tilde{\mathscr{C}}=\{X|X=\nu(R)U_R^{\dagger}YU_R, Y\in \mathscr{C}, \nu:\mathbb{R}^{n_r}\to \mathbb{C}\}$ to be a set of commutative random observables. The map $\tilde{\mathbb{P}}(\cdot |\tilde{\mathscr{C}})$ is called (a version of) the quantum-classical conditional expectation from $\tilde{\mathscr{C}}'$ onto $\tilde{\mathscr{C}}$, if $\tilde{\mathbb{P}}(\tilde{\mathbb{P}}(X |\tilde{\mathscr{C}})Y)=\tilde{\mathbb{P}}(XY)$ for all $X\in \tilde{\mathscr{C}}'$ and $Y\in \tilde{\mathscr{C}}$.}

\textcolor{blue}{\textbf{Theorem A1.} \cite{Gao2015} (Quantum-classical Bayes formula) Consider the classical probability space model $(\Omega,\mathcal{F},\mathcal{P})$, the set of random observables $\mathscr{C}$ and the quantum-classical expectation operator $\tilde{\mathbb{P}}$ defined in Section III. Suppose a new probability measure $\mathcal{Q}$ is defined by $d\mathcal{Q}=\Lambda d\mathcal{P}$, where the $\mathcal{F}-$measurable random variable $\Lambda$ is the classical Radon-Nikon derivative. Choose $V\in \tilde{\mathscr{C}}'$ such that $V^{\dagger}V>0$ and $\tilde{\mathbb{P}}(\Lambda V^{\dagger}V)=1$. Then we can define on $\tilde{\mathscr{C}}'$ a new quantum-classical expectation operator $\tilde{\mathbb{Q}}$ by $\tilde{\mathbb{Q}}(X)=\tilde{\mathbb{P}}(\Lambda V^{\dagger}XV)$ and
\begin{equation}
\tilde{\mathbb{Q}}(X|\tilde{\mathscr{C}})=\frac{\tilde{\mathbb{P}}(\Lambda V^{\dagger}XV/\tilde{\mathscr{C}})}{\tilde{\mathbb{P}}(\Lambda V^{\dagger}V/\tilde{\mathscr{C}})}, \hspace{1cm}\forall X\in \tilde{\mathscr{C}}'. \label{gquantum9}
\end{equation}}

\emph{Proof of Lemma 3.1.} Let $\tilde{\mathbb{Q}}^t$ be a normal state as $\tilde{\mathbb{Q}}^t(X)=\tilde{\mathbb{P}}(U^{\dagger}(t)XU(t))$.

Note that $\mathscr{Y}_t=U^{\dagger}(t)\mathscr{Q}_tU(t)$ follows from the fact that $U^{\dagger}(t)Q(s)U(t)=U^{\dagger}(s)Q(s)U(s)$. From Definition A1 one can obtain that
\begin{eqnarray}
\tilde{\mathbb{P}}(\left<F(t),e_j\right>U^{\dagger}(t)XU(t)|\mathscr{Y}_t)=U^{\dagger}(t)\tilde{\mathbb{Q}}^t(\left<F(t),e_j\right>X|\mathscr{Q}_t)U(t) \label{app2}
\end{eqnarray}
almost surely under $\tilde{\mathbb{P}}$.

In addition, suppose the system is initialized at $\pi_0=\sum \limits_{k} p_k \left|\alpha_k\right>\left<\alpha_k\right|$ and we define a curve $\left|\psi_k(t)\right>=U(t)(\left|\alpha_k\right>\otimes \left|\upsilon \right>)$. Using the fact that $dB(t)\left|\upsilon \right>=0$,  one obtains (see Equation (6.13) in \cite{Holevo1991})
\begin{equation}
d\left|\psi_k(t)\right>=\{(-iH(F(t))-\frac{1}{2}L^{\dagger}L)dt+LdQ(t)\}\left|\psi_k(t)\right>. \label{app3}
\end{equation}
In other words, $U(t)(\left|\alpha_k\right>\otimes \left|\upsilon \right>)=V(t)(\left|\alpha_k\right>\otimes \left|\upsilon \right>)$ since $U(0)=V(0)=I$. After some mathematical manipulation, one obtains $\Tr(\rho_0 U^{\dagger}(t)XU(t))=\Tr(\rho_0 V^{\dagger}(t)XV(t))$ which leads to
\begin{equation}
\tilde{\mathbb{P}}(\left<F(t),e_j\right>U^{\dagger}(t)XU(t))=\tilde{\mathbb{P}}(\left<F(t),e_j\right>V^{\dagger}(t)XV(t)). \label{app4}
\end{equation}
Then we can apply Theorem A1 by replacing $(\Lambda, X, V, \tilde{\mathscr{C}})$ by $(1, \left<F(t),e_j\right>X, V(t), \mathscr{Q}_t)$ and obtain
\begin{eqnarray}
\tilde{\mathbb{Q}}^t(\left<F(t),e_j\right>X|\mathscr{Q}_t)=\frac{\tilde{\mathbb{P}}(\left<F(t),e_j\right> V^{\dagger}(t)XV(t)|\mathscr{Q}_t)}{\tilde{\mathbb{P}}( V^{\dagger}(t)V(t)|\mathscr{Q}_t)}.  \label{app5}
\end{eqnarray}
Lemma 3.1 can be concluded combining (\ref{app2}) and (\ref{app5}).

\end{document}